# Quantum Dots at Room Temperature carved out from Few-Layer Graphene

*Amelia Barreiro[1, 2, *], Herre S. J. van der Zant[1], Lieven M. K. Vandersypen[1]*

[1] Kavli Institute of Nanoscience, Delft University of Technology, Lorentzweg 1, 2628 CJ Delft, The Netherlands.

[2] Department of Physics, Columbia University, New York, New York 10027, USA.

AUTHOR EMAIL ADDRESS ab3690@columbia.edu

ABSTRACT. We present graphene quantum dots endowed with addition energies as large as 1.6 eV, fabricated by the controlled rupture of a graphene sheet subjected to a large electron current in air. The size of the quantum dot islands is estimated to be in the 1 nm range. The large addition energies allow for Coulomb blockade at room temperature, with possible application to single-electron devices.

MANUSCRIPT TEXT. Graphene is a one-atom-thick planar sheet of sp$^2$-bonded carbon atoms whose shape can be structured by means of standard top-down fabrication techniques, representing a simple and scalable approach to realize electronic devices. Indeed, one of the most appealing research directions involving graphene is its use as the base material for electronic circuitry that is envisaged to consist of nanometer-sized elements.[1,2] For this purpose, graphene nanoribbons have captured widespread attention.[3-11] Also quantum dot (QD) devices made entirely from graphene are considered, with possible applications to single-electron transistors and supersensitive electrometry.[12,13]

Most QDs reported to date operate at cryogenic temperatures, which limits their use in applications. In practice, the two most important conditions for room-temperature operation are addition energies much larger than the thermal energy at 300 K, i.e. >> 26 meV, and stable device operation. Individual molecules in between two electrical contacts can act as quantum dots and addition energies are usually in excess of 100 meV.[14-16] Nevertheless, the operation of molecular devices at room temperature is often limited due to the high atomic mobility of the metallic electrodes which renders them unstable at room temperature.[17] Recently, it was reported that single P atoms in a Si lattice act as QDs, but room-temperature operation remains to be demonstrated.[18] Room-temperature QDs and single-electron transistors have been realized in ultra-small metallic grains,[19] and in etched silicon devices[20,21]. Furthermore, RT-QDs have been achieved within an individual metallic carbon nanotube (CNT), by introducing a kink in individual CNTs by atomic force microscopy (AFM) manipulation.[22] However, it was not possible to completely suppress the current at room temperature due to thermal smearing. Furthermore, these CNT RT-QDs are not easy to fabricate and cannot easily be scaled up to form QD arrays.

Here, we report a simple method to fabricate all-graphene quantum dots that can operate at room temperature. These graphene QDs are endowed with addition energies ($E_{add}$) that can be as large as 1.6 eV. Their formation relies on the controlled rupture of a few-layer graphene sheet subjected to a large electron current. Simple estimates show that the size of the charge-carrier island of these quantum dots lies in the 1 nm range.

We start by briefly describing our fabrication technique. Few-layer graphene flakes (between 3 - 18 nm thick) are deposited by mechanical exfoliation of kish graphite (Toshiba Ceramics) on degenerately doped silicon substrates coated with 280 nm of thermal silicon oxide. For the exfoliation, we use standard wafer protection tape as it leaves little adhesive residue on substrates. Electrodes are patterned on top of selected few-layer graphene flakes by electron-beam lithography and subsequent Cr/Au evaporation, followed by lift-off in cold acetone and dichloroethane. Initial device resistances at low bias are in the order of 200 Ω - 3 kΩ.

We now proceed to the formation of quantum dot devices by electroburning in air,[23-25] using a similar technique to the one reported for current-annealing of graphene,[26,27] or the controlled rupture of shells of multi-wall carbon nanotubes.[28-30] Typically, a voltage ($V$) ramp is applied to the few-layer graphene flake (1 V/s), while the current ($I$) is continuously recorded with 200 μs sampling intervals. The variations in the conductance ($G = I/V$) are monitored, with a feedback condition set at a >10 % drop in $G$ within the past 200 mV of the ramp. Upon the occurrence of such a drop, the voltage is swept back to zero in 100 μs. Immediately after, a new sweep starts from zero voltage and the process is repeated, in this way gradually narrowing down the flake (see figure S1 in the supporting information). The process can be repeated until a nanometer spaced gap is formed.[24] Just before the formation of a nano-gap, a very narrow connection is left between the two bigger parts of the flake,[25] forming either a single quantum dot or several quantum dots in series (see figure S2 in the Supporting Information). Figure 1 shows a schematic of a nanometer-sized quantum dot carved out of a graphene sheet by electroburning and an AFM image of a typical device.

When the two-point resistance at zero gate voltage ($V_g$) at 100 mV source-drain voltage ($V$) exceeds 100 MΩ, the feedback controlled electroburning is stopped and stability diagrams as a function of $V_g$ and $V$ are taken to determine if several dots in series or a single dot were formed (more details are given below). If several quantum dots in series are found, the electro-burning procedure can be applied once again in order to obtain a single dot. The fabrication technique has a reasonable yield: out of 36 devices studied, 9 exhibited single quantum dot behaviour (see table S1 for their properties), 11 showed signatures of QDs coupled in series, and 13 were burned through completely, forming open gaps between the two electrodes. See table S1 in the Supporting Information for details of the properties of the fabricated single-dot devices.

In the devices exhibiting quantum-dot behavior, we find that the conductance $G$ of the graphene sheet at low source-drain bias is a strong function of the back gate voltage ($V_g$), showing Coulomb peaks in some cases. Measuring the current ($I$) as a function of source-drain bias ($V$) and $V_g$ can yield well defined Coulomb diamonds for some devices, which indicate the presence of charge carrier islands in the ultra-narrow constriction. Figures 2 and 3 (and figures S3 and S4 in the supporting information) display typical examples of such Coulomb diamonds, measured in different samples at $T$ =10 K. In Fig. 2a, the diamond is closed, which is usually a sign that a single island is formed. Note that, unfortunately, we were not able to record multiple Coulomb diamonds on most samples, because it was not possible to sweep the gate to larger positive or negative values due to the limit set by the electrical breakdown of our $SiO_2$ gate dielectric (typically ≈ 60 V). Nevertheless, for some samples we were able to resolve more than two halves of the Coulomb diamonds with similar heights (see figure 4), thus demonstrating that not narrow constrictions but quantum dots were formed. Indeed, a nanoribbon would give rise to a real bandgap,[32-34] in which case a single region of suppressed current is expected.

In the Coulomb blockade regime, the diamond´s height along the $V$-axis is a measure of the energy, $E_{add}$, needed to add one charge carrier to the island. The addition energies we measure have strikingly large values, up to 1.6 eV, see figure 2a. The value of 1.6 eV is approx. one order of magnitude larger than $E_{add}$ in the largest addition energy graphene single electron transistors reported so far.[13,23]

Apart from the addition energy, figure 2 gives information on the quantized level spacing ($\Delta E$), the energy difference between consecutive discrete orbitals on the islands. When not only the ground state but also an excited state electrochemical potential falls inside the bias window, there are two transport channels through the dot instead of just one. This increases the probability for electrons to pass through the island and gives rise to a stepwise increase of the current (see figure 2). In figure 2a, $\Delta E$ is approximately 0.8 eV, assuming that no excitations occur below 0.4 eV. From $\Delta E$ and $E_{add}$, the charging energy can be extracted, $2E_c = e^2/C_T = E_{add} - \Delta E \sim 0.8$ eV, where $C_T$ is the total capacitance of the dot. In contrast to earlier reports,[13] we thus find that $E_c$ cannot be neglected even for ultrasmall graphene islands.

It is surprising that without intentionally introducing tunnelling barriers between the graphene electrodes and the charge carrier island, confinement strong enough for quantum dot formation takes place. In patterned graphene quantum dots, two constrictions in series are defined lithographically, creating a small island in between.[13,38] In our case, it is unlikely that two similar or even narrower constrictions forming an island in between have formed unintentionally, nor has it been observed on suspended devices formed by a similar mechanism.[23,25] Quantum dot formation has been observed in graphene nanoribbons on a $SiO_2$ substrate and was ascribed to a quantum confinement energy gap combined with edge disorder and charge inhomogeneities in the $SiO_2$ which give rise to charge puddles.[38-40] We do not believe our quantum dots result from charge inhomogeneities in the substrate because quantum dot formation by electroburning was also observed in suspended samples.[23] In the latter case, island formation was ascribed to charge carriers becoming localized by potential fluctuations along the ultra-narrow constriction in the presence of a confinement gap due to e.g. molecules having reacted with the dangling bonds.[23] More generally, any type of edge disorder could be responsible for localization of charges and island formation.

We now turn to estimating the size of our smallest quantum dot, namely the one measured in figure 2a. The level spacing $\Delta E$ is approximately 0.8 eV and can be used to estimate the island size. Using the band structure of monolayer graphene and a square confinement potential, the diameter of the carrier island can be estimated according to $d = \pi \hbar v_F / \Delta E \sim 2.6$ nm. Here, $v_F \sim 10^6$ m/s is the Fermi velocity and $\hbar = 6.582 \cdot 10^{-16}$ eV·s is the reduced Planck constant. Since we cannot exclude that the dot is formed from bilayer graphene, we also estimate the size of the dot using the band structure of bilayer graphene and a square potential and obtain $d = \sqrt{\frac{\hbar^2 \pi}{m^* \cdot \Delta E}} \sim 3 nm$, where $m^* \sim 0.033 m_e$ is the effective mass in bilayer graphene ($m_e$ is the electron mass),[41] and obtain a similar value as for monolayer graphene.

A second independent estimate of the size of the graphene charge carrier island can be obtained from the charging energy of the dot $E_c = e^2/2(C_G+C_D+C_S+C_{self}) \sim 0.4$ eV, where $C_S$ and $C_D$ are the capacitances to the source and drain, respectively, and $C_{self}$ is the self-capacitance of the carrier island of the quantum dot. From the stability diagram in figure 2a we can extract the values $C_G = 1.66 \cdot 10^{-21}$ F, $C_S = 4.15 \cdot 10^{-20}$ F and $C_D = 5.76 \cdot 10^{-20}$ F.[27] Calculating $C_T = C_G+C_D+C_S+C_{self} = e^2 / 2E_C \sim 2 \cdot 10^{-19}$ F we see that the total capacitance is mostly given by $C_{self}$. We approximate $C_T \sim C_{self}$ to obtain a further estimate of the size of the carrier island by modelling it as a circular disk of diameter d, for which $d = e^2 / 4 \cdot \varepsilon_0 \cdot (\varepsilon_r+1) \cdot E_c$, with $\varepsilon_0$ the vacuum permittivity and $\varepsilon_r = 3.9$ the relative dielectric constant of $SiO_2$. This approach gives $d = 2.26$ nm.

Another rough estimate of the island size can be obtained from the back gate capacitance. We can estimate the dot diameter from $C_g$ by looking at two extreme scenarios. One model is that of a parallel plate capacitor and the other one corresponds to a small disk above an infinite plane (no screening from the electrodes). Using the formula for a parallel plate capacitor $C_g = \varepsilon_0 \cdot \varepsilon_r \cdot \pi \cdot (d/2)^2/D$, where $D = 285$ nm is the thickness of the SiO$_2$, we obtain $d \sim 4$ nm. For a small disk on an infinite plane, the capacitance is given by $C_g = \varepsilon_0 \cdot (1+\varepsilon_r) \cdot 2d$,[42] resulting in $d \sim 0.018$ nm. Due to screening by the electrodes, most of the back gate area does not contribute to the capacitance to the island so the latter estimate is far off. Taking for illustration purposes the capacitance of the island to a 7 nm diameter disc located 285 nm underneath, the island size would again be $d \sim 1$ nm.

Overall, the measurements suggest that the diameter of the island of the quantum dot displayed in figure 2a is on the 1 nm scale.

To put these results in perspective, the charging energy $E_c \sim 0.4$ eV in even our smallest quantum dot is almost one order of magnitude bigger than the $E_c \sim 50$ meV of a quantum dot consisting of a single phosphorus atom in a Si substrate, which corresponds to the binding energy of an isolated phosphorus donor in bulk silicon.[18,43] As a matter of comparison, the diameter of a phosphorus dopant ($d_P$) in Si corresponds to twice its effective Bohr radius ($r_{eB}$) which can be estimated according to $d_P = 2r_{eB} = 4.4$ nm,[44] in this case with a spherical symmetry instead of a disk shape.

Importantly, we are able to measure Coulomb blockade even at room temperature, see figures 3 and 4, and not only single *I-V* traces, but also stability diagrams which are stable over large ranges in $V_g$ and $V$. The device in figure 3 appears completely insulating with no measurable conductance over an extended range of $V_g$ (>50 V). Room-temperature operation is possible because $E_c$ and $\Delta E$, and consequently also $E_{add}$, in our devices are much bigger than the thermal energy $k_B T$, which at $T$=300 K corresponds to 25.6 meV. Moreover, the stability of graphene at elevated temperatures due to its strong sp$^2$ bonds is another important factor to enable room temperature operation of our quantum dots.

Nevertheless, in all the quantum dots measured at room temperature, more noise is present in the data as compared to the measurements taken at low temperature, see figures 3 and 4. Apart from that, sudden jumps in $V_g$ sometimes occur while taking full stability diagrams, as illustrated in figure 4. These limitations in device performance are likely to be set by nearby dopants and trapped charges, for example in the silicon oxide, that are mobile at room temperature. We believe that these factors, extrinsic to the graphene island, are the likely origin of the noise and scatter present in figure 3 and 4.[44,45] Indeed, the motion of a single dopant could alter the dielectric environment and therefore the current through the QDs. Another possible source of noise could be the thermally activated motion of polar dangling bonds in the vicinity of the quantum dot which could possibly also modify the dielectric environment.[46] Also the switching events such as the one occurring in figure 4 at $V_g \approx$ -45 V are likely to be caused by trapped charges moving in the dielectric silicon oxide layer. The trapped charges are more mobile at room temperature and can result in sudden jumps, causing switches of the electrostatic environment, which in turn lead to sudden jumps in the current through the graphene quantum dot. Indeed, the presence of just one dopant can alter the performance of short-channel transistors depending on where they are located.[44] Possibly a gate dielectric with only few trapped charges, such as hBN,[47] could be used to reliably operate these devices without random changes in the dielectric environment, thereby reducing the noise and switching events present in the current data.

In conclusion, we have shown that the rupture of a graphene sheet subjected to a large current can be harnessed to fabricate graphene quantum dots where the active component is very close to the ultimate physical limit of Moore's law. The as-fabricated quantum dots are endowed with addition energies as large as 1.6 eV. We estimate the size of the carrier island of the quantum dots to lie in the 1 nm range. Remarkably, graphene remains stable and conductive at the nanometer scale and the observed large addition energies give rise to Coulomb blockade at room temperature, a prerequisite for most applications. We further note that the fabrication technique is not limited to the use of exfoliated graphene but could also be applied to CVD-grown few-layer graphene over large areas, paving the path to more complex, integrated devices involving multiple QD devices integrated on the same chip.

ACKNOWLEDGMENT. We gratefully acknowledge discussions with P. Kim, F. Prins, S. Nadj-Perge and A. van der Zande. Financial support was obtained from the Dutch Foundation for Fundamental Research on Matter (FOM) and the Agència de Gestió d´Ajuts Universitaris i de Recerca de la Generalitat de Catalunya (2010_BP_A_00301).

SUPPORTING INFORMATION PARAGRAPH. **Supporting Information Available**. The Supporting Information contains further details about the electroburning procedure, a table with the properties of all the fabricated single dot devices, data of quantum dots in series, further measurements of the devices in figure 2 b, as well as data of other large addition energy quantum dots both at low and at room temperature.

FIGURE CAPTIONS.

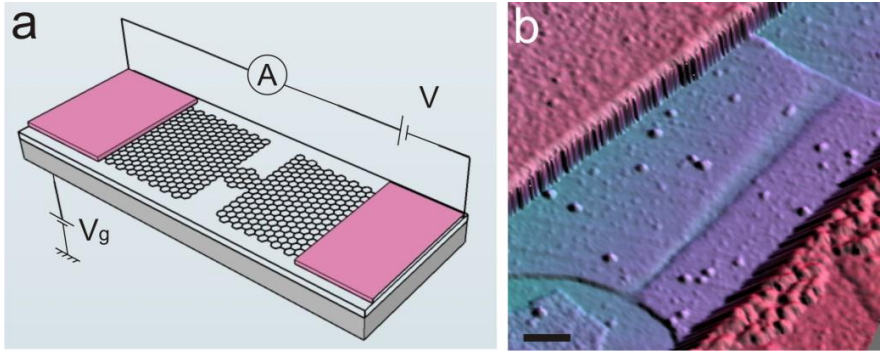

**Figure 1.** (a) Schematic of the device and measurement set-up. (b) AFM image of a typical device.[31] A gap is formed via electroburning, separating the few-layer graphene flake into two parts which can be connected through a connection so tiny that it can´t be imaged by AFM. Remarkably, this tiny connection can form a single quantum dot with addition energies as large as 1.6 eV. The scale bar is 400 nm.

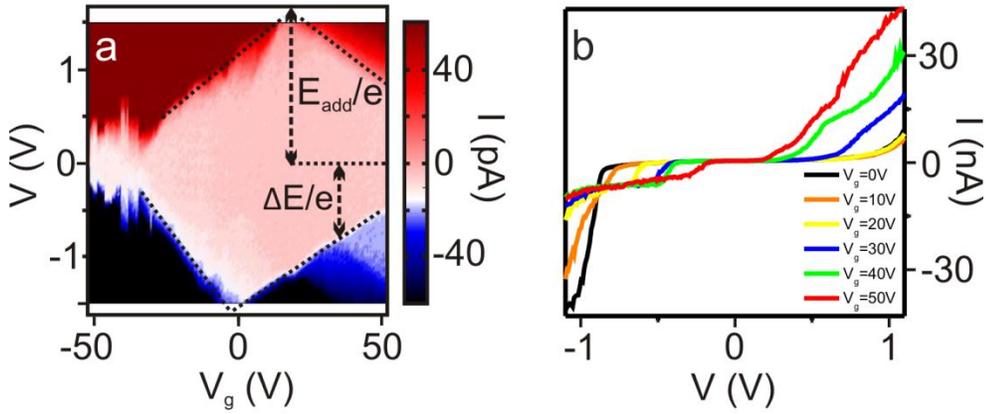

**Figure 2.** Large addition energy quantum dots at 10 K. (a) Current map of sample 1 in table S1 as a function of the applied bias voltage $V_b$ and gate voltage $V_g$ from which $E_{add} \sim 1.6$ eV can be extracted. The noise centered at $V_g \sim -40$ V could possibly have resulted from an environmental charge instability or could also have resulted from a second (larger) disorder-induced island. On the lower right side a stepwise increase of the current due to an excited state entering the bias window can be observed. A quantized level spacing of $\Delta E \sim 0.8$ eV can be extracted. The fact that the excited level is only visible on one side of the Coulomb diamond (in this case at negative voltage) can be attributed to an asymmetry in the coupling to the leads.[35,36] (b) Current-voltage traces at various $V_g$ for device 2 in table S1. The stepwise increase of the current is due to an excited state entering the bias window, thereby increasing the probability for electrons to pass through the island. The full stability diagram of this device is shown in Fig. S3.[37]

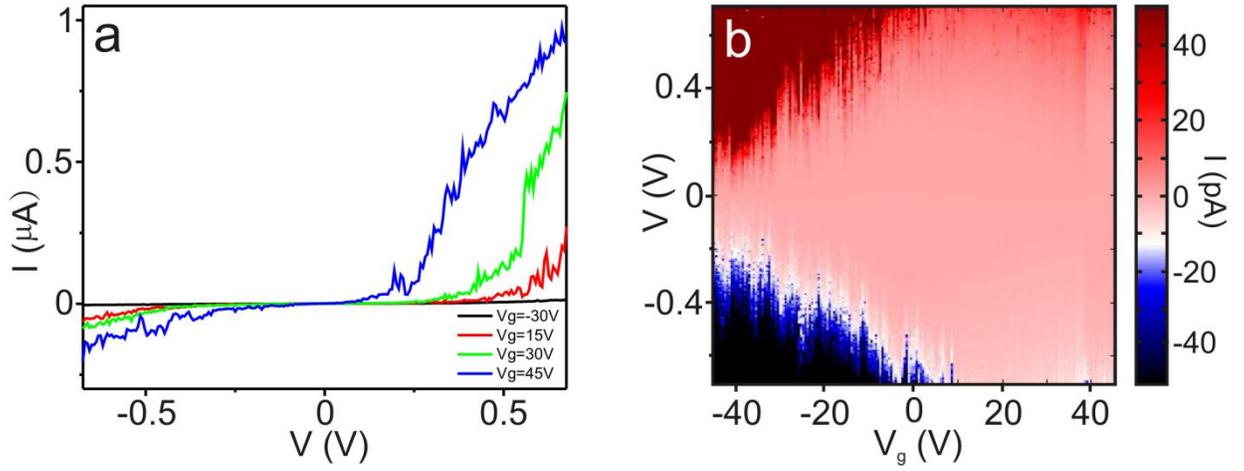

**Figure 3.** Room-temperature operation of graphene quantum dots (device 5 in table S1).[37] (a) *I-V* traces for four different gate voltages. (b) Current map as a function of $V_g$ and *V*; the conductance is fully suppressed over a large gate-voltage range (over 50 V).

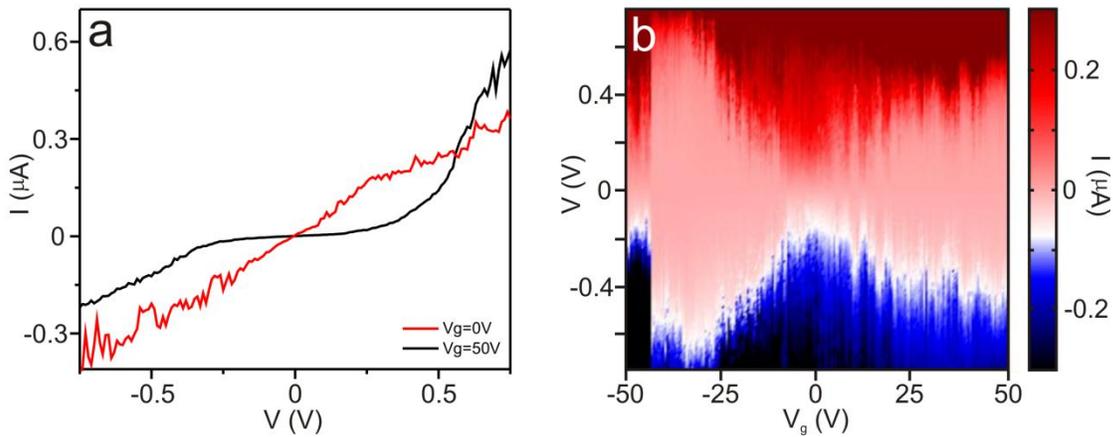

**Figure 4.** Room temperature operation of quantum dots (sample 4 in table S1).[37] (a) I-V traces taken at Vg = 0 V (red) and Vg = 50 V (black). (b) Current map as a function of Vg and V at room temperature. Two diamond-shaped regions of suppressed current can be observed. The jump at Vg=-45V probably originates from an environmental charge instability.

REFERENCES.

(1) Geim, A. K.; Novoselov, K. S.; Nat. Mater. **2007,** *6*, 183-191.

(2) Castro Neto, A. H.; Guinea, F.; Peres, N. M. R.; Novoselov, K. S.; Geim, A. K. Rev. Mod. Phys. **2009**, *81*, 109–162.

(3) Han, M. Y.; Ozyilmaz, B.; Zhang, Y. B.; Kim, P. Phys. Rev. Lett. **2007**, *98*, 206805.

(4) Avouris, P. ; Chen, Z. H. ; Perebeinos, V. Nat. Nanotechnol. **2007**, *2*, 605.

(5) Son, Y. W. ; Cohen, M. L.; Louie, M S. G.; Nature **2006**, *444*, 347.

(6) Gunlycke, D.; Areshkin, D. A.; White, C. T.; Appl. Phys. Lett. **2007**, *90*, 142104.

(7) Yang, L.; Park, C. H.; Son, Y. W.; Cohen, M. L. ; Louie, S. G. Phys. Rev. Lett. **2007**, *99*, 186801.

(8) Peres, N. M. R.; Castro Neto, A. H. ; Guinea, F. Phys. Rev. B **2006**, *73*, 195411.

(9) Barone, V.; Hod, O.; Scuseria, G. E. ; Nano Lett. **2006**, *6*, 2748.

(10) Brey, L.; Fertig, H. A. Phys. Rev. B **2006**, *73*, 235411.

(11) Wunsch, B.; Stauber, T.; Guinea, F. Phys. Rev. B **2008**, *77*, 035316.

(12) Likharev, K. K. Proc. of the IEEE **1999**, *87*, 606.

(13) Ponomarenko, L. A.; Schedin, F.; Katsnelson, M. I.; Yang, R.; Hill, E. W.; Novoselov, K. S.; Geim, A. K. Science **2008**, *320*, 356.

(14) Kubatkin, S.; Danilov, A.; Hjort, M.; Cornil, J.; Brédas, J.-L.; StuhrHansen, N.; Hedegård, P.; Bjørnholm, T. Nature **2003**, *425*, 698.

(15) Osorio, E. A.; O'Neill, K.; Stuhr-Hansen, N.; Nielsen, O. F.; Bjørnholm, T.; van der Zant, H. S. J. Adv. Mater. **2007**, *19*, 281.

(16) Ward, D. R.; Scott, G. D.; Keane, Z. K.; Halas, N. J.; Natelson, D. J. Phys.: Condens. Matter **2008**, *20*, 374118.

(17) Strachan, D. R.; Smith, D. E.; Fischbein, M. D.; Johnston, D. E.; Guiton, B. S.; Drndic, M.; Bonnel, D. A.; Johnson, A. T. Nano Lett. **2006**, *6*, 441-444.

(18) Fuechsle, M.; Miwa, J. A.; Mahapatra, S.; Ryu, H.; Lee, S.; Warschkow, O.; Hollenberg, L. C. L.; Klimeck, G.; Simmons, M. Y. Nat. Nanotech. **2012**, *7*, 242–246.

(19) Matsumoto, K.; Ishii, M.; Segawa, K.; Oka, Y.; Vartanian, B. J.; Harris, J. S. Appl. Phys. Lett. **1996**, *68*, 34.

(20) Takahashi, Y.; Naease, M.; Namatsu, H.; Kurihara, K.; Iwdate, K.; NakajiGa, Y.; Horiguchi, S.; Murase, K.; Tabe, M. Electron. Lett. **1995**, *31*, 136-137.

(21) Zhuang, L.; Guo, L.; Chou, S. Y. Appl. Phys. Lett **1998**, *72*, 1205.

(22) Postma, H. W. Ch.; Teepen, T.; Yao, Z.; Grifoni, M.; Dekker, C. Science **2001**, *293*, 76-79.

(23) Moser, J.; Bachtold, A. Appl. Phys. Lett. **2009**, *95*, 173506.

(24) Prins, F.; Barreiro, A.; Ruitenberg, J. W.; Seldenthuis, J. S.; Aliaga-Alcalde, N.; Vandersypen, L. M. K.; van der Zant, H. S. J. Nanoletters **2011**, *11*, 4607 - 4611.

(25) Barreiro, A.; Börrnert, F.; Rümmeli, M. H.; Büchner, B.; Vandersypen, L. M. K. Nanoletters **2012**, *12*, 1873-1878.

(26) Moser, J.; Barreiro, A.; Bachtold, A. Appl. Phys. Lett. **2007**, *91*, 163513.

(27) Barreiro, A.; Rurali, R.; Hernández, E. R.; Bachtold, A. Small **2011**, *7*, 775-780.

(28) Collins, P. C.; Arnold, M. S.; Avouris, P. Science **2001**, *292*, 706–709.

(29) Bourlon, B.; Glattli, D. C.; Plaçais, B.; Berroir, J. M.; Miko, C.; Forró, L.; Bachtold, A. Phys. Rev. Lett. **2004**, *92*, 026804.

(30) Barreiro, A.; Rurali, R.; Hernandez, E. R.; Moser, J.; Pichler, T.; Forro, L.; Bachtold, A. Science **2008**, *320*, 775–778.

(31) Horcas, I.; Fernández, R.; Gómez-Rodríguez, J. M.; Colchero, J.; Gómez-Herrero, J.; Baro, A. M. Rev. Sci. Instrum. **2007**, *78*, 013705.

(32) Han, M. Y.; Oezyilmaz, B.; Zhang, Y.; Kim, P. Phys. Rev. Lett. **2007**, *98*, 206805.

(33) Son, Y.-W., Cohen, M. L., Louie S. G. Phys. Rev. Lett. **2006**, *97*, 216803.

(34) Wang, X.; Ouyang, Y.; Li, X.; Wang, H.; Guo, J.; Dai, H. Phys. Rev. Lett. **2008**, *100*, 206803.

(35) van Houten, H.; Beenakker, C. W. J.; Staring, A. A. M. Single Charge Tunneling edited by H. Grabert and M. H. Devoret, NATO ASI Series B294 (plenum, New York, 1992).

(36) Thijssen, J. M.; van der Zant.; H. S. J. Phys. Stat. Sol B **2008**, *245*, 1455-1470.

(37) Please see the Supporting Information.

(38) Stampfer, C.; Güttinger, J.; Hellmüller, S.; Molitor, F.; Ensslin, K.; Ihn, T. Phys Rev. Lett. **2009**, *102*, 056403.

(39) Todd, K.; Chou, H.-T.; Amasha, S.; Goldhaber-Gordon, D. Nano Letters **2009**, *9* 416-421.

(40) Gallagher, P.; Todd, K.; Goldhaber-Gordon, D. Phys. Rev. B **2010**, *81*, 115409.

(41) Adam, S.; Das Sarma, S. Phys. Rev. B **2008**, *77*, 115436.

(42) H. A. Wheeler, IEEE Transactions on Microwave Theory and Techniques **1982**, *30*, 2050-2054.

(43) Ramdas, A. K.; Rodriguez, S. Rep. Prog. Phys. **1981**, *44*, 1297–1387.

(44) Pierre, M. ; Wacquez, R. ; Jehl, X. ; Sanquer, M. ; Vinet, M. ; Cueto, O. Nature Nanotech. **2010**, *5*, 133-137.

(45) Rogge, S. Nat. Nanotech. **2010**, *5*, 100-101.

(46) Guo, X.; Small, J. P.; Klare, J. E.; Wang, Y.; Purewal, M.; Hong, B. H.; Caldwell, R.; Huang, L.; O´Brien, S.; Yan, J.; Wind, S. J.; Hone, J.; Kim, P.; Nuckols, C. Science **2006**, *311*, 356-359.

(47) Dean, C. R.; Young, A. F.; Meric, I.; Lee, C.; Wang, L.; Sorgenfrei, S.; Watanabe, K.; Taniguchi, T.; Kim, P.;  Shepard, K. L.; Hone, J. Nat. Nanotech. **2010**, *5*, 722-726.

TOC.

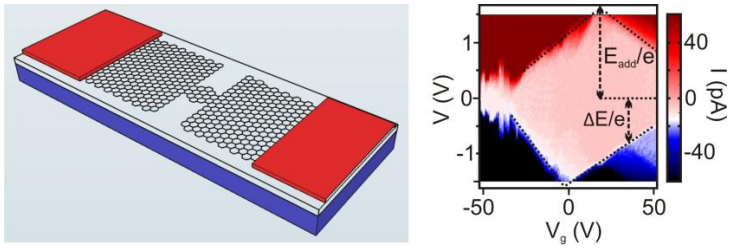

# Supplementary Information

# Quantum Dots at Room Temperature carved out from Few-Layer Graphene

*Amelia Barreiro[1, 2, \*], Herre S. J. van der Zant[1], Lieven M. K. Vandersypen[1]*

[1] Kavli Institute of Nanoscience, Delft University of Technology, Lorentzweg 1, 2628 CJ Delft, The Netherlands.

[2] Department of Physics, Columbia University, New York, New York 10027, USA.

AUTHOR EMAIL ADDRESS ab3690@columbia.edu

# S1. Electroburning

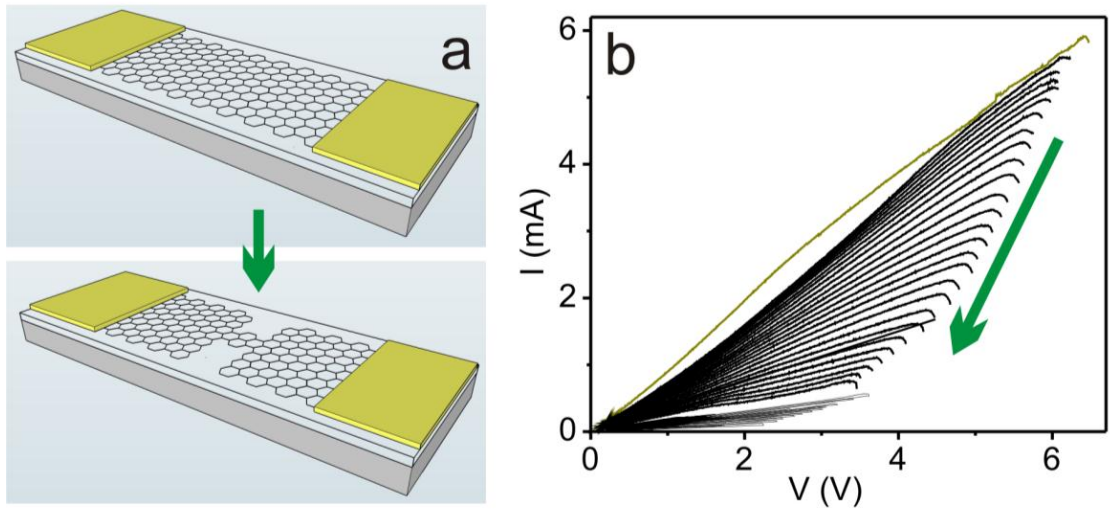

**Figure S1.** (a) Schematic of the feedback-controlled electroburning process, before (top) and after (bottom), the formation of an ultrasmall graphene island. (b) Current-voltage (I-V) traces of the evolution (green arrow) of the feedback-controlled electro-burning. The first I-V trace is displayed in green.

Feedback controlled electroburning is performed in air at room temperature and is based on a similar procedure as used for the electromigration of metallic nanowires [1, 2]. Here, the source-drain voltage ($V$) is ramped up at a speed of approx. 1 V/s, while the current ($I$) is continuously recorded with a 200 μs sampling rate. The variation in the conductance ($G = I/V$) is monitored with a feedback condition set at a >10% drop in G within the past 200 mV of the ramp. Upon the occurrence of such a drop, the voltage is rapidly swept back to zero in 100 μs. Immediately after, a new sweep starts from zero voltage. The process is repeated until the few-layer graphene flake has narrowed down into a nanometer-sized island. Figure S1b shows a typical evolution of feedback-controlled electroburning. Generally, during the first voltage ramp (green trace in Figure S1b) nonlinear *I-V* characteristics are observed, probably due to removal of contaminants on the flake by current-induced annealing [3]. Increasing the voltage further induces the first electroburning event, as can be seen from the downward curvature in the *I-V* characteristic, in this case at V = 6.4 V and I = 5.89 mA. The feedback then sweeps the voltage back to 0 V and a new voltage ramp is started. As the electroburning process evolves, the conductance decreases in steps and the voltage at which the electroburning occurs decreases (see green arrow in Figure S1b). When the two-point resistance at zero gate voltage ($V_g$) at V=100 mV exceeds 100 MΩ, stability diagrams as a function of $V_g$ and $V$ are taken. If several quantum dots in series are found, the electro-burning procedure can be applied once again in order to obtain a single dot. This fabrication procedure has a reasonable yield: out of 36 devices studied, 9 exhibited single quantum dot behaviour, 11 showed signatures of QDs coupled in series, and 13 were burned through completely, forming open gaps between the two electrodes.

| sample | temperature | $E_{add}$ | $\Delta E$ | $E_c$ |
|---|---|---|---|---|
| 1 | 10K | 1.6 eV | 0.8 eV | 0.4 eV |
| 2 | 10K | 1.1 eV | 0.65 eV | 0.225 eV |
| 3 | 10K | 1.3 eV | No info | No info |
| 4 | Room T. | 0.7 eV | No info | No info |
| 5 | Room T. | >0.7 eV | No info | No info |
| 6 | 10K | 0.6 eV | No info | No info |
| 7 | Room T. | 0.5 eV | No info | No info |
| 8 | 10K | >0.7 eV | No info | No info |
| 9 | 10K | >1 eV | No info | No info |

**Table S1. Properties of the nine fabricated single-dot devices and the temperature at which the measurements were performed.**

## S2. Quantum dots in series

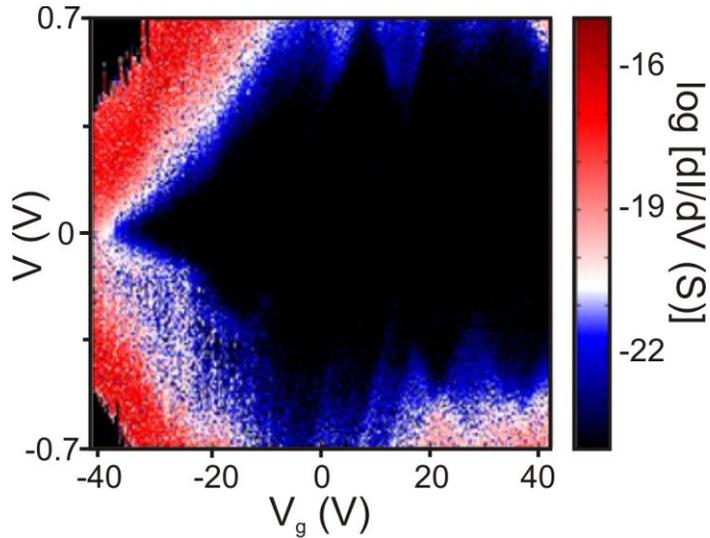

**Fig. S2. Differential conductance *dI/dV* plotted in log-scale as a function of $V_g$ and *V* measured at 10 K displaying a Coulomb blockade pattern arising from several islands in series.**

## S3. Additional data of the device in figure 2b

From the stability diagram in figure S4b the following energy scales are extracted:

$E_{add} \sim 1.1$ eV, $\Delta E \sim 0.65$ eV. With these values we estimate $E_c = (E_{add} - \Delta E) / 2 = 0.225$ eV.

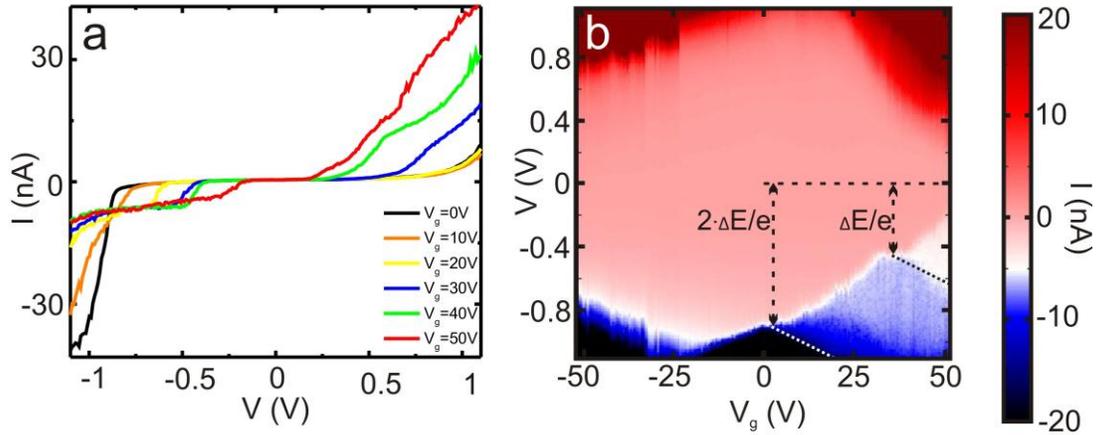

**Figure S3.** Device charaterization at 10 K. (a) I-V traces for different $V_g$. (b) Current map as a function of $V_g$ (sample 2 in table S1).

## S4. Other examples of large addition energy quantum dots

From the stability diagram in figure S5c the following energy scales were extracted for this device:

$E_{add} \sim 1.3$ eV.

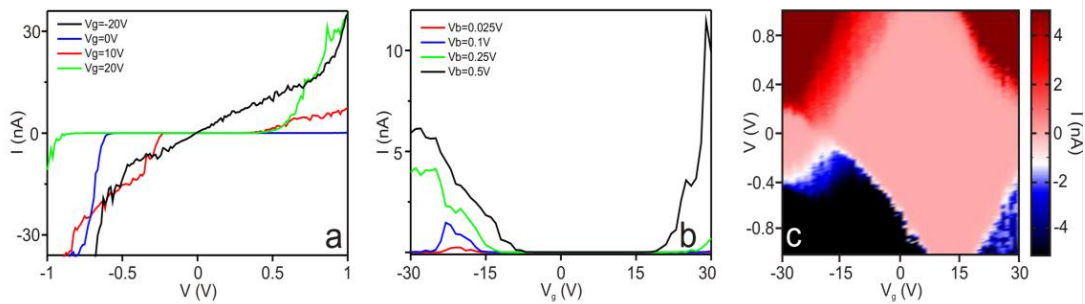

**Figure S4.** Device charaterization at 10 K of sample 3 in table S1. (a) I-V traces at different $V_g$. (b) Current as a function of $V_g$ for different bias voltages. (c) Current map as a function of $V_g$ and $V$ at 10 K.

**References:**

1. Prins, F.; Hayashi, T.; van Steenwijk, B. J. A. D.; Gao, B.; Osorio, E. A.; Muraki, K.; van der Zant, H. S. J. Appl. Phys. Lett. 2009, 94, 123108.

2. Strachan, D. R.; Smith, D. E.; Johnston, D. E.; Park, T. H.; Therien, M. J.; Bonnell, D. A.; Johnson, A. T. Appl. Phys. Lett. 2005, 86, 043109.

3. Moser, J.; Barreiro, A.; Bachtold, A. Appl. Phys. Lett. 2007, 91, 163513.